\documentclass[12pt]{iopart}

\usepackage{iopams}
\usepackage{textcomp}
\usepackage{graphicx}
\usepackage[colorlinks=true, linkcolor=blue, citecolor=blue, urlcolor=blue]{hyperref}
\usepackage{cite}
\usepackage{caption}
\usepackage{textcomp}
\begin{document}

\title{Endcap-Type Paul Trap for Precision Spectroscopy and Studies of Controlled Interactions}

\author{Anand Prakash$^{1,*}$, Akhil Ayyadevara$^1$, E. Krishnakumar$^1$, M. Ibrahim$^1$, K. M. Yatheendran$^1$, Subhadeep De$^2$, Sayan Patra$^{1}$\footnote{Present address (Sayan Patra):  Lawrence Livermore National Laboratory, 7000 East Ave, Livermore, CA 94550, USA}, S. A. Rangwala$^{1,\dagger}$}

\address{$^1$Raman Research Institute, C. V. Raman Avenue, Sadashivanagar, Bangalore 560080, India}
\address{$^2$The Inter-University Centre for Astronomy and Astrophysics (IUCAA), Postbag 4, Ganeshkhind, Pune 411007, Maharashtra, India}

\address{$^{*,\dagger}$Authors to whom correspondance should be addressed}
\ead{prakash@rri.res.in, sarangwala@rri.res.in}

\vspace{10pt}

\begin{abstract}
We present the design and fabrication of an endcap-type Paul trap. The trap is designed for studies with Ca$^{+}$ and Yb$^{+}$.  The design, fabrication process, and characterization are presented in detail with a focus on trapping a single compensated ion at the rf node. A custom-built imaging system of $NA = 0.14$ and magnification $\approx 22 \times$ performs close to diffraction-limit and resolves multi-ion clusters. Controlled ion loading and characterization of the trap are performed using $^{40}$Ca$^{+}$. The experimentally determined quadrupole coefficient of the trap is $\approx 0.3$, which is very close to the design value. The relative frequency shift along the spectroscopy beam due to excess micromotion (EMM) is at the level of $3.5\times 10^{-18}$ for $^{40}$Ca$^{+}$. Applications of this trap encompass single-ion-based optical frequency standards, tests of fundamental physics, the study of mesoscopic Coulomb clusters, and the controlled interaction of a single ion with co-trapped atoms. 
\end{abstract}

\section{Introduction}
Atomic and molecular ions trapped in a radio-frequency (rf) Paul trap can be isolated from the environment and laser-cooled, enabling better control over the system's state, longer trapping lifetimes, reduced systematic effects, and longer coherence times. These properties make it one of the ideal platforms for precision spectroscopy and atomic clocks \cite{diddamsOpticalClockBased2001a,rosenbandFrequencyRatioHg2008,chouFrequencyComparisonTwo2010,huntemannHighAccuracyOpticalClock2012,barwoodAgreementTwo882014,dubeSr88Singleion2015,huntemannSingleIonAtomicClock2016a,brewer27QuantumLogicClock2019a,kellerProbingTimeDilation2019a,arnoldPrecisionMeasurementsBa2020,kingOpticalAtomicClock2022,zengTransportableCaOptical2023,zhiqiang176LuClock2023a,tofful171YbOptical2024a,marshallHighStabilitySingleIonClock2025,hausser115Yb1722025,lindvall88SrOptical2025a,yuNuclearSpinQuenching2025b,marceauAbsoluteFrequencyMeasurement2025}, tests of fundamental physics \cite{berengutOpticalTransitionsHighly2012,godunFrequencyRatioTwo2014b,pruttivarasinMichelsonMorleyAnalogue2015,safronovaSearchNewPhysics2018,mehlstaublerAtomicClocksGeodesy2018,megidishImprovedTestLocal2019,countsEvidenceNonlinearIsotope2020,hurEvidenceTwoSourceKing2022a,barontiniMeasuringStabilityFundamental2022,dreissenImprovedBoundsLorentz2022a,sherrillAnalysisAtomicclockData2023,filzingerImprovedLimitsCoupling2023a,filzingerUltralightDarkMatter2025,doorProbingNewBosons2025}, quantum computation and simulations \cite{porrasEffectiveQuantumSpin2004,blattEntangledStatesTrapped2008,haffnerQuantumComputingTrapped2008a,friedenauerSimulatingQuantumMagnet2008,blattQuantumSimulationsTrapped2012,barreiroOpensystemQuantumSimulator2011,aharonyshapiraInverseMpembaEffect2024}, study of phase transitions in mesoscopic systems \cite{diedrichObservationPhaseTransition1987,dubinNormalModesCold1996,ulmObservationKibbleZurek2013,yanExploringStructuralPhase2016,kietheFinitetemperatureSpectrumSymmetrybreaking2021,ducaOrientationalMeltingMesoscopic2023,ruffertDomainFormationStructural2024,ayyadevaraObservingDynamicsOctupolar2025a, ayyadevaraSymmetrycontrolledThermalActivation2026} and trapped ion-atom interactions \cite{grierObservationColdCollisions2009,zipkesTrappedSingleIon2010,hallLightAssistedIonNeutralReactive2011,raviCoolingStabilizationCollisions2012,harterSingleIonThreeBody2012,hazeObservationElasticCollisions2013}. 

The endcap trap \cite{schramaNovelMiniatureIon1993a} is a geometrical variation of the classical ring-type Paul trap \cite{paulNotizenNeuesMassenspektrometer1953}, where the ring electrode is replaced by coaxial shields placed around the endcap electrodes (see Figure \ref{fig:endcaptrap}). This modification provides better optical access for cooling, probe beams, and imaging systems, albeit at a reduced quadrupole efficiency. Due to a single rf null point in the trapping region, its application is mainly limited to single-ion precision spectroscopy \cite{huntemannHighAccuracyOpticalClock2012,huntemannSingleIonAtomicClock2016a,lindvall88SrOptical2025a,huangCa40Ion2019a} and tests of fundamental physics \cite{filzingerImprovedLimitsCoupling2023a,filzingerUltralightDarkMatter2025}. However, the scope of this trap geometry can be extended to attempt qubit operations in 2D and 3D crystals in the presence of micromotion \cite{donofrioRadialTwoDimensionalIon2021a,wuHighfidelityEntanglingGates2021}, study of Coulomb crystals in a 3D trap \cite{diedrichObservationPhaseTransition1987,ayyadevaraObservingDynamicsOctupolar2025a, ayyadevaraSymmetrycontrolledThermalActivation2026}, interaction between a single ion and a cloud of trapped atoms, and portable optical clocks \cite{spampinatoIonTrapDesign2024a}.

In this article, we describe the design details of the complete apparatus, including the parameters considered for the trap electrode design and assembly process, the vacuum system, the rf drive circuit, and the imaging system. By loading and exciting the notion of a single ion, the trap is characterized against its design parameters. Next, micromotion compensation is performed in 3D, and the relative frequency due to residual micromotion along the spectroscopy beam is evaluated. We then demonstrate the performance of a custom-built lens system by resolving the structural transitions of multi-ion clusters.  The characterization results of the apparatus make it suitable for performing precision spectroscopy and studying controlled interactions using Coulomb crystals.

\section{Ion Trap Apparatus}
\subsection{Trap Design}
\begin{figure}
\centering
 \includegraphics[width=\linewidth]{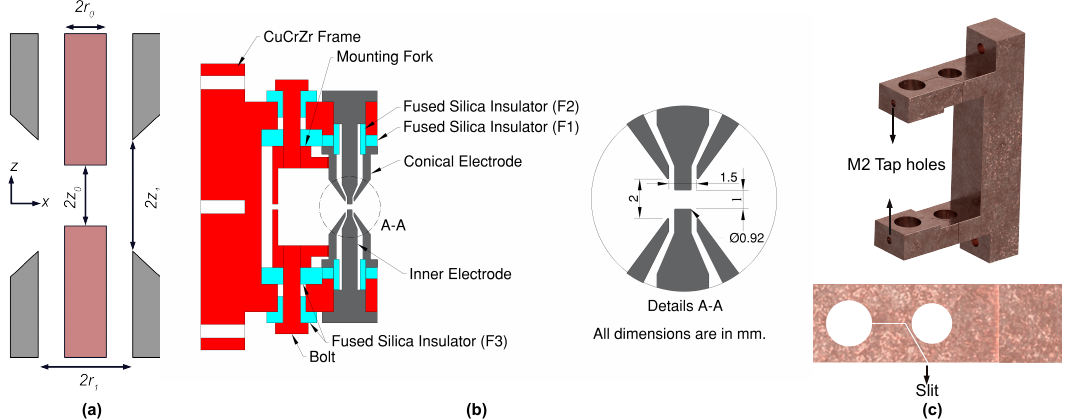}
\caption{(a) A $2D$ schematic diagram of the trap electrodes showing the relevant design parameters. (b) Central plane view of the designed trap. (c) (Top) M2 tap holes on one face, (Bottom) Slit on the perpendicular face to align the inner electrodes.}

\label{fig:endcaptrap}
\end{figure}

The endcap-type trap \cite{schramaNovelMiniatureIon1993a} consists of a pair of coaxial cylindrical electrodes (inner electrodes) surrounded by a pair of concentric conical outer electrodes, as shown in Figure \ref{fig:endcaptrap}. Inner electrodes carry the rf voltage, and the outer conical electrodes are at ground potential or biased with a dc offset.

Critical electrode dimensions, shown in Figure \ref{fig:endcaptrap}, $2r_{0}$, $2z_{1}$, $2r_{1}$, and $2z_{0}$ are decided by a combination of optimization of the quadrupole coefficient, minimization of the contribution from higher-order pole terms, reduction of deposition of loading material on the inner electrodes, and machining viability of electrode dimensions. Since the Full-Width Half Maximum of the atomic beam at the trap centre is $0.96$ mm \cite{prakashLowDivergenceColdwall2024a}, we fix $2z_{0} = 1$ mm (see Figure \ref{fig:endcaptrap}) to avoid coating of the inner electrodes in the long run. The distance between the inner and the outer electrode $(2r_{1})$ is optimized to minimize the trap capacitance and optimize the quadrupole coefficient. The value of the quadrupole coefficient is maximum for $2z_{1} = 2z_{0}$. However, it comes at the cost of a high contribution from higher-order multipole terms \cite{spampinatoIonTrapDesign2024a}. The separation between conical electrodes ($2z_{1}$) is increased until the change in the higher-order multipole contribution is sufficiently small with further increase. 

A C-shaped frame is used to secure the inner electrodes in position. An L-shaped fork is used to mount the conical electrode shield. The rf voltage to the trap is provided via the C-shaped frame. The L-shaped fork shields the effect of the mounting frame at the trap centre \cite{nisbet-jonesSingleionTrapMinimized2016a}. Thus, only the electrode geometry is considered for simulating potential and analyzing the contribution of multipoles. An electrostatic simulation is performed in COMSOL\textsuperscript{\textregistered} $5.6$ to generate the potential in the trapping region. Considering the rotational symmetry about the $z-$axis and the mirror symmetry about the $z=0$ plane present in our system (check Figure \ref{fig:endcaptrap}), the potential generated in the trapping region will be of the following form 
\begin{eqnarray}
\phi(r,z) &=& A_0
+ \frac{A_{2}}{2z_{0}^{2}}(2z^{2} - r^{2})
+ \frac{A_{4}}{8z_{0}^{4}}(8z^{4} + 3r^{4} - 24r^{2}z^{2}) \nonumber \\
&& + \frac{A_{6}}{16z_{0}^{6}}(16z^{6} - 5r^{6}
- 120r^{2}z^{4} + 90r^{4}z^{2}) \nonumber \\
&& + \frac{A_{8}}{128z_{0}^{8}}(128z^{8} + 35r^{8}
- 1792z^{6}r^{2} + 3360z^{4}r^{4} - 1120z^{2}r^{6}) \nonumber \\
&& + \frac{A_{10}}{256z_{0}^{10}}(256z^{10} - 63r^{10}
- 5760z^{8}r^{2} + 20160z^{6}r^{4} \nonumber \\
&& - 16800z^{4}r^{6} - 3150z^{2}r^{8}), \label{eq:trapping_potential}
\end{eqnarray}
where  $r^{2} = x^{2}+y^{2}$, $A_{2}$ is the quadrupole coefficient and $A_{4}$, $A_{6}$, $A_{8}$, and $A_{10}$ are higher-orderr multipole coefficients. The value of the fitted quadrupole coefficient ($A_{2}$) for our designed trap is $\approx 0.3$; $A_{4}/A_{2} \approx 0.008$, $A_{6}/A_{2} \approx 0.089$, $A_{8}/A_{2} \approx 0.009$, and $A_{10}/A_{2}\approx 0.013$.

\subsection{Material Selection}
Ideally, the trap's metal parts should be easy to machine and have high electrical conductivity, low thermal emissivity, and high mechanical strength. High electrical conductivity decreases Joule heating, and low emissivity reduces the blackbody radiation shift. Molybdenum (Mo), tungsten (W), tantalum (Ta), copper (Cu), and copper chromium zirconium (CuCrZr) were the high-purity materials accessible. Among these available materials, and considering the above properties, Mo was chosen for the electrodes, and CuCrZr was chosen for the frame and other connecting parts. The surface of the inner electrode close to the ion is polished to reduce emissivity.

The power loss per unit volume in a dielectric placed in an ac field \cite{kumphElectricfieldNoiseThin2016} is directly proportional to the frequency of the electric field, $f$,  the relative permittivity of the dielectric, $\epsilon_{r}$, $\tan(\delta)$, where $\delta$ dielectric loss tangent, and square of the electric field amplitude, $|E| ^{2}$. The high relative permittivity of insulators increases the capacitance of the trap system. Insulators are made as thin as possible to decrease the system's capacitance. High relative permittivity results in a higher current through the metal parts, thus increasing Joule heating. Fused silica and sapphire are the most preferred insulator materials; in our case, fused silica is used as the insulator material. A detailed comparison of the properties of metal and insulator parts is provided in Dole\v{z}al \textit{et al} \cite{dolezalAnalysisThermalRadiation2015a}. 

\subsection{Trap Assembly}
The trap frame guides all the alignments. The coaxiality of the inner electrodes is maintained by the frame that holds the electrodes. Transition fit is used to position the inner electrodes in the frame. An M2 tap hole and a slit on the face perpendicular to the tap hole face are made for fine adjustment of the coaxiality, as shown in Figure \ref{fig:endcaptrap}. The slit acts as a flexure spring, holding the electrode in its final position. To maintain concentricity, fused silica spacers are used to position the outer electrode with respect to the inner electrode. Fused silica or Macor\textsuperscript{\textregistered} insulators isolate the outer electrodes from the trap frame in the non-critical parts. The electrode dimensions are inspected using a coordinate measuring machine, and the measured dimensions are compared with the design dimensions in Table \ref{electrode-dimensions}. The electrode surfaces exposed to the ion are characterised using an atomic force microscope (AFM).

\begin{table}
\caption{Design and machined dimensions of the trap electrodes}
\label{electrode-dimensions}
\footnotesize
    \begin{tabular}{@{}p{4cm}p{2.5cm}p{2.5cm}p{2.5cm}p{2.5cm}@{}}
    \br
    Parameters & $2r_{0}$ (mm)& $2r_{1}$ (mm) & $(z_{1}-z_{0})$  (mm) & $2z_{0}$ (mm)\\
    \mr       
    Design Value & $0.920$ & $1.50$ & $0.500$ & $1.00$\\
    Machined Dimensions & $0.905$, $0.908$ & $1.48$, $1.51$ & $0.460$, $0.468$ & $1.03$\\
    \br 
    \end{tabular}\\
\end{table}

\begin{figure}
    \centering
    \includegraphics[width=\linewidth]{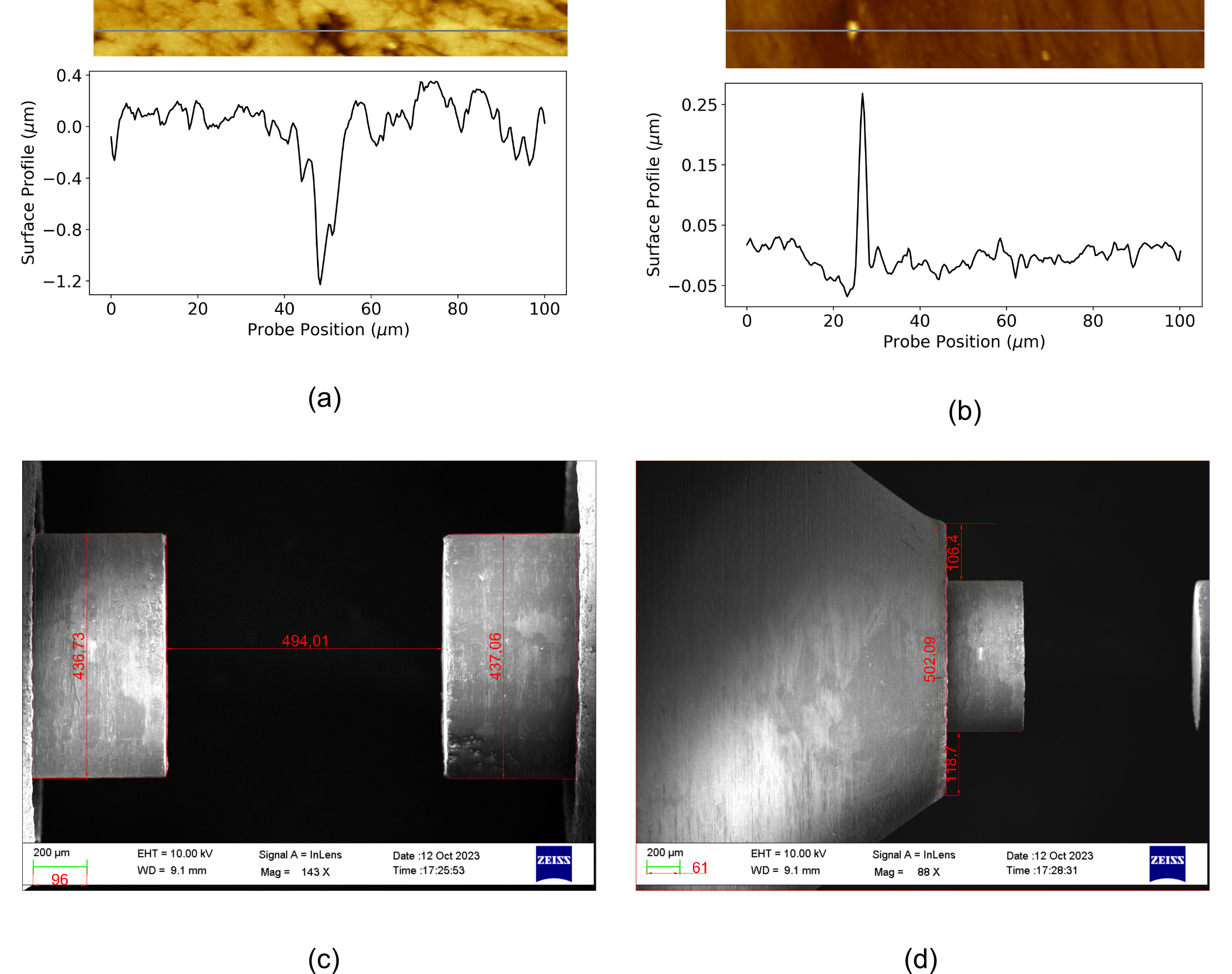}
    \caption{(a) (Top) AFM image of a selected region of the first inner electrode. (Bottom) Surface profile along the line shown in the top image.  (b) (Top) AFM image of a selected region of the second inner electrode. (Bottom) Surface profile along the line shown in the top image. (c) SEM image showing dimensions of the inner electrode and the distance between them in the aligned assembly. (d) SEM image showing the alignment of the outer electrode with respect to the inner electrode in one direction of the trap assembly.}
    \label{fig:AFM-SEM}
\end{figure}

\normalsize
The root mean square roughness of the roughest patch ($100$ $\mu$m $\times$ $100$ $\mu$m) of one electrode is 208 nm, and that of the other electrode is 26 nm. AFM image of both electrode surfaces along a line having the worst patch is shown in Figure \ref{fig:AFM-SEM}. Molybdenum is a porous material, making it challenging to achieve a reproducible smooth surface finish through grinding. Mechanical polishing is performed to reduce the surface roughness. Although it has improved the surface quality, we still observed a few micron-sized undulations, as shown in Figure \ref{fig:AFM-SEM}.
The complete trap assembly is first aligned under a high-magnification imaging system to adjust the coaxiality of the electrodes. When the best alignment is achieved, scanning electron microscope (SEM) imaging is performed to quantify it.

A small quantity of Torr Seal\textsuperscript{\textregistered} is applied at the edge of the mating surfaces of the inner electrodes and the trap frame to retain the alignment during the baking process. SEM images of the aligned trap assembly are shown in Figure \ref{fig:AFM-SEM}.
\begin{figure}
    \centering
    \includegraphics[width=\linewidth]{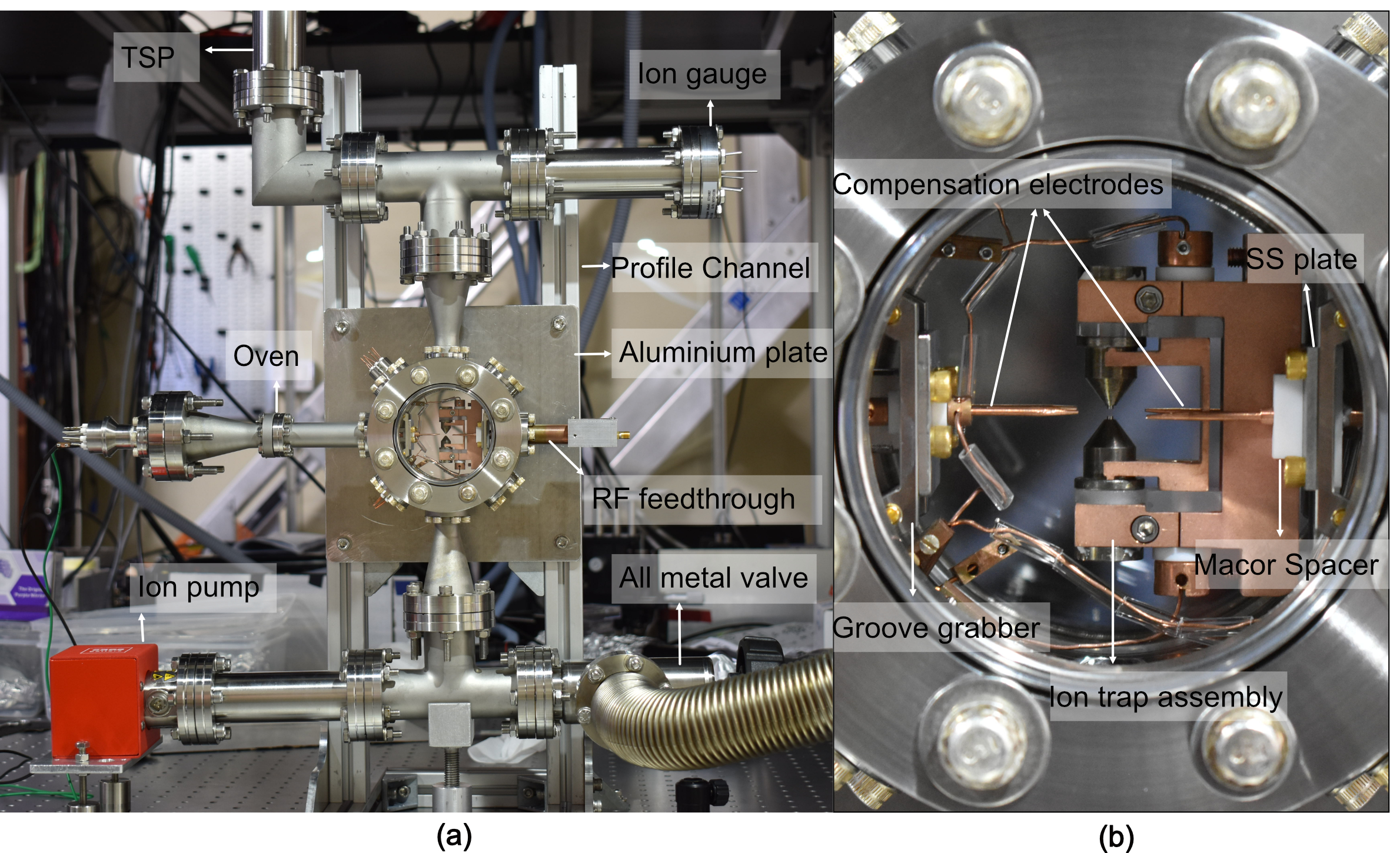}
    \caption{(a) Complete vacuum setup. (b) A close view of the trap assembly with connecting wires mounted inside the chamber.}
    \label{fig:complete-assembly}
\end{figure}
The trap is mounted inside a spherical octagon chamber from Kimball Physics, model no: \textit{MCF450-SphOct-E2A8}, with the help of groove grabbers. A high-voltage feedthrough from VACOM, CF16-HV20S-1-CE-CU24, provides the rf connection to the trap body. High transmission Corning HPFS 7980 windows are used for sending beams and collecting fluorescence. A titanium sublimation pump, along with a non-evaporable getter ion pump, NEXTORR 100-D, is used to achieve the final pressure of $\sim 2\times 10^{-10}$ mbar. Figure \ref{fig:complete-assembly} shows the trap mounted inside the vacuum chamber, including the compensation electrodes used to compensate for excess micromotion (EMM) in the radial plane. The EMM along the axial direction is compensated using the conical electrodes.

\subsection{Drive Circuit}
\begin{figure}
    \centering
    \includegraphics[width=\linewidth]{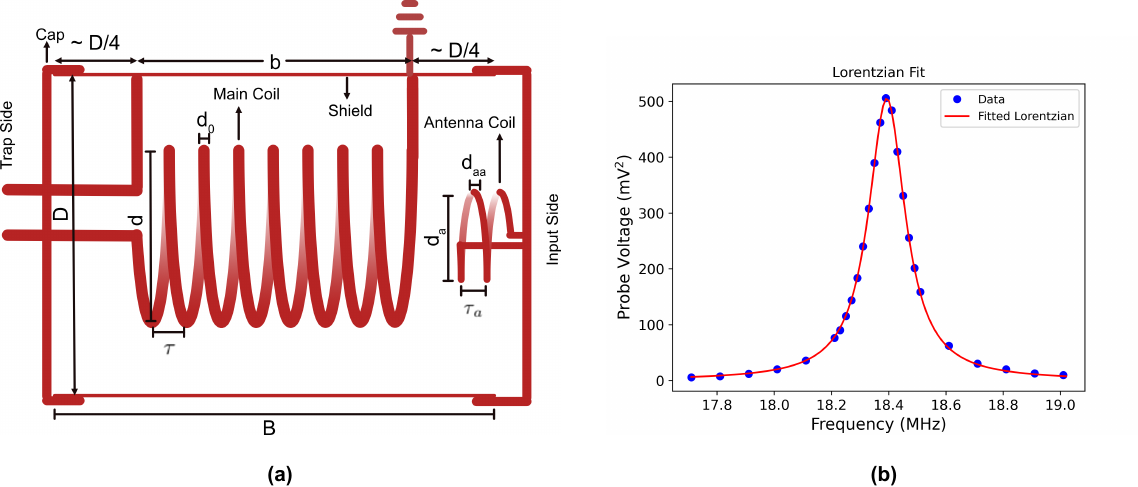}
    \caption{(a) Schematic diagram of the helical resonator. Parameters of the resonator are: d $= 53$ mm, D $= 90$ mm, d$_{0} = 6$ mm,  b $ = 80$ mm, $\tau = 10$ mm, d$_{a} = 31$ mm, $\tau_{a} = 5$ mm, d$_{aa} = 2$ mm. The number of turns in the main coil and the antenna coils are 8 and 2, respectively. (b) Voltage measured across the probe at different drive frequencies.}
    \label{fig:Resonator-loaded}
\end{figure}

A helical resonator drives the trap in the frequency range of $18-19$ MHz. It helps in matching impedance for efficient transfer of the applied power to the trap. Impedance is matched by changing the distance between the main coil and the antenna coil \cite{panjaNoteMeasuringCapacitance2015a}. The schematic of the resonator is shown in Figure \ref{fig:Resonator-loaded}. The voltage applied across the trap is measured using a voltage probe \cite{parkNewMeasurementMethod2021b} with an effective capacitance of 1 pF, added in parallel to the trap. The voltage probe has two capacitors, 1 pF and 1 nF, in series. The voltage applied to the trap is sampled across the 1 nF capacitor in the $\approx 1/1000$ ratio. It is also used to characterize the resonator; the transmitted voltage across the probe is shown in Figure \ref{fig:Resonator-loaded}. In the loaded case, $Q$ is $\approx 115$, and the resonant frequency is $\approx 18.4$ MHz. An rf choke, consisting of two $22 \; \mu$H inductors in series, is added after the resonator to apply the dc offsets to the trap. After the addition of the choke, the resonance frequency of the trap changed to $18.26$ MHz. All compensation and ground electrodes are capacitively grounded using a parallel combination of $0.1$ pF and $1$ pF capacitors to discharge the pickup voltages.

\subsection{Imaging System}
\begin{figure}
    \centering
    \includegraphics[width=\linewidth]{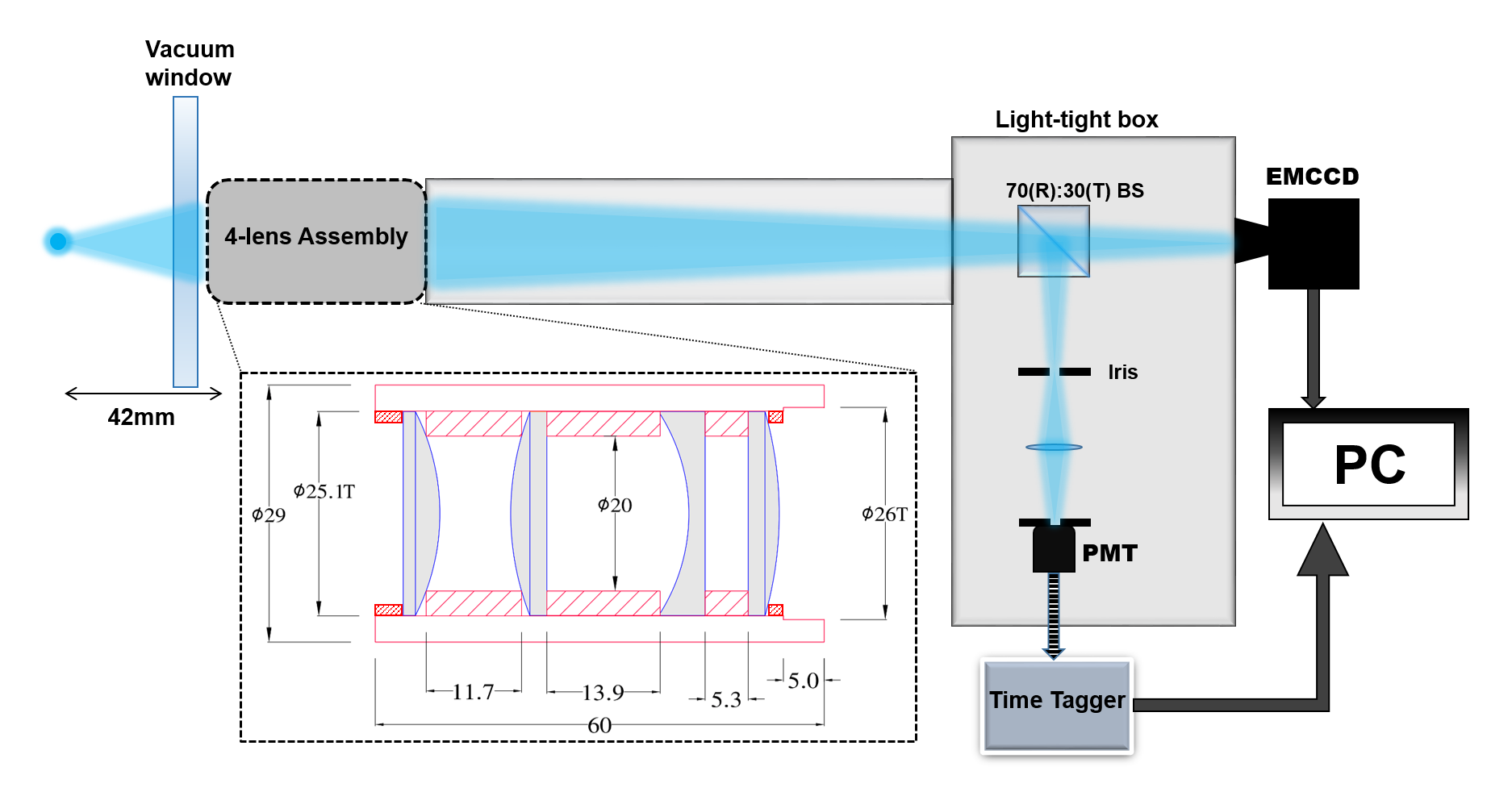}
    \caption{Schematic diagram of the fluorescence imaging system. The 4-lens objective is housed in a home-built brass tube with brass spacers. The objective guides the fluorescence into a light-tight box, where a 70:30 beam splitter transmits 30$\%$ of the fluorescence onto an EMCCD camera and reflects 70$\%$ of the fluorescence onto a PMT. A spatial filter before the PMT ensures a high signal-to-noise ratio (SNR) for photon counting. }
    \label{fig:img-sys}
\end{figure}

The trapped $^{40}$Ca$^+$ ions are detected by the fluorescence emitted at 397 nm. The imaging objective should have a high numerical aperture to collect scattered photons for photon counting with a photo-multiplier tube (PMT). The objective must also be diffraction-limited to focus the fluorescence of a single ion for efficient spatial filtering to achieve high SNR. We have designed a 4-lens objective with commercial AR-coated fused silica lenses with high transmittance in the UV range. The separation between each lens is optimized with ZEMAX to achieve a peak wavefront error of $<\frac{\lambda}{25}$. The distance from the trap center to the vacuum viewport limits the working distance to $> 42$ mm. The design NA is 0.14, and the magnification is $22 \times$. The schematic of the fluorescence detection system is shown in Figure \ref{fig:img-sys}. A $70:30$ beam splitter is placed at the end of the imaging system to simultaneously observe the signal on a single photon counting PMT from Hamamatsu, H10682-210, and an EMCCD Andor iXon 897.

\section{Trap Performance and Characterization of Apparatus}
\subsection{Ion Loading and Detection}
\begin{figure}
    \centering
    \includegraphics[width=0.7\linewidth]{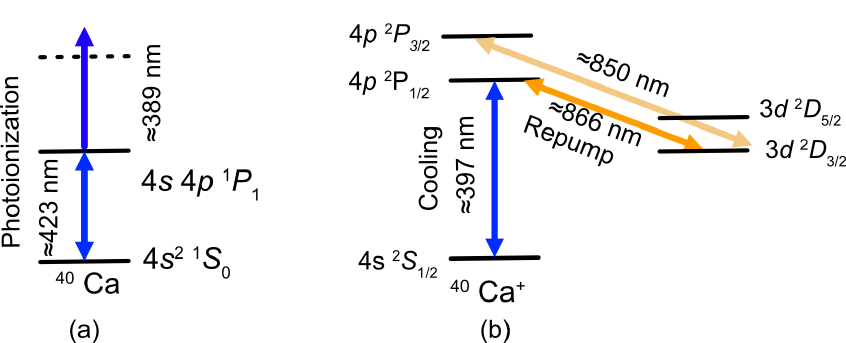}
    \caption{(a) Ionization scheme of $^{40}$Ca atom (b) Cooling and repump transitions of $^{40}$Ca$^{+}$ ion, 850 nm laser is used to excite $4p$ $^{2}P_{3/2}$ transition.}
    \label{fig:Cooling-ion cloud}
\end{figure}

The presented experimental setup is characterized using calcium ions. Two-photon ionization scheme, as shown in Figure \ref{fig:Cooling-ion cloud}, is used to create ions in the trapping region. The first laser beam drives the $4s^{2}$ $^{1}S_{0} \rightarrow 4s4p$ $^{1}P_{1}$ transition of calcium atoms, and the second beam takes it to the ionization continuum from the $4s4p$ $^{1}P_{1}$ state. The second beam is derived from a fibre-coupled Thorlabs LED, model no: MF365FP1. The 850 nm laser is a home-built external cavity diode laser; the remaining lasers are from Toptica.

\begin{figure}
    \centering
    \includegraphics[width=\linewidth]{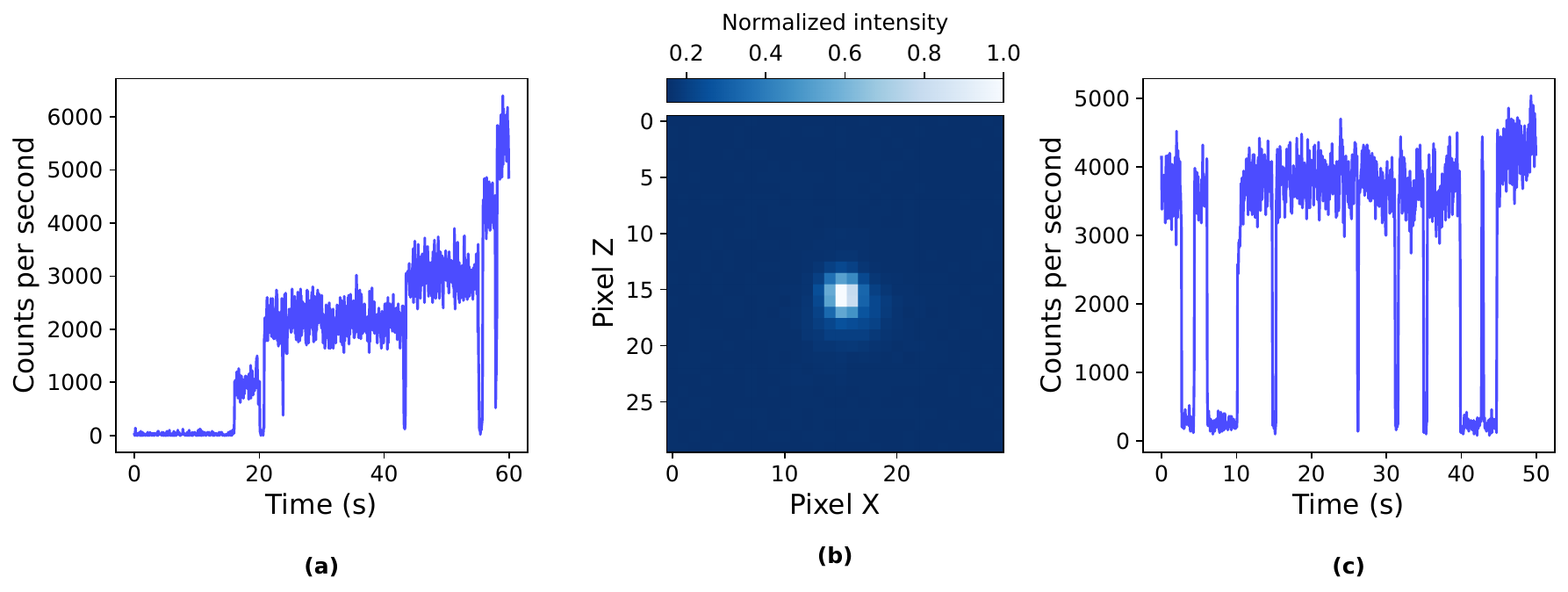}
    \caption{(a) Ion loading signal observed on the PMT (b) False colour image of a single $^{40}$Ca$^{+}$ ion. (c) Quantum jumps for a single $^{40}$Ca$^{+}$ ion observed on PMT.  }
    \label{fig:ion-loading}
\end{figure}

For controlled loading of ions, oven current is set to $4.1$ A. Power in the first and the second ionization beam is set to 80 $\mu$W and $\leq$ 15 $\mu$W. The detuning of the cooling beam is set to $-60$ MHz, and the repump beam is tuned to the resonance. The ion loading process initiates after switching on the LED, and trapping of each ion is registered as a jump in the fluorescence signal observed on the PMT. The electron shelving method \cite{dehmelt1975} can also be used to count the number of ions loaded in the trap. A laser at 850 nm is added along with the cooling and the repump laser to address $3d\  ^{2}D_{3/2} \rightarrow 4p\ ^{2}P_{3/2}$ transition. Power in the 850 nm beam is $\approx$ 4 nW. The $4p\ ^{2}P_{3/2}$ state decays to $4s\ ^{2}S_{1/2}$ and $3d\ ^{2}D_{5/2}$ with branching ratio of $\approx 17:1$ \cite{gerritsmaPrecisionMeasurementBranching2008}. The $3d\ ^{2}D_{5/2}$ is a metastable state having a lifetime of $1168(9)$ ms \cite{kreuterExperimentalTheoreticalStudy2005}. When the population reaches the metastable state, the ions become dark. In the case of a single ion, the PMT signal ``jumps" from the signal to the background level (quantum jumps). For two ions, either it goes to the half or to the background level, depending on the number of ions in the dark state. This technique enables us to verify the number of ions loaded in the trap. The quantum jump signal for a single ion is shown in Figure \ref{fig:ion-loading}. 

\subsection{Determination of Quadrupole Coefficient}

\begin{figure}
    \centering
    \includegraphics[width=\linewidth]{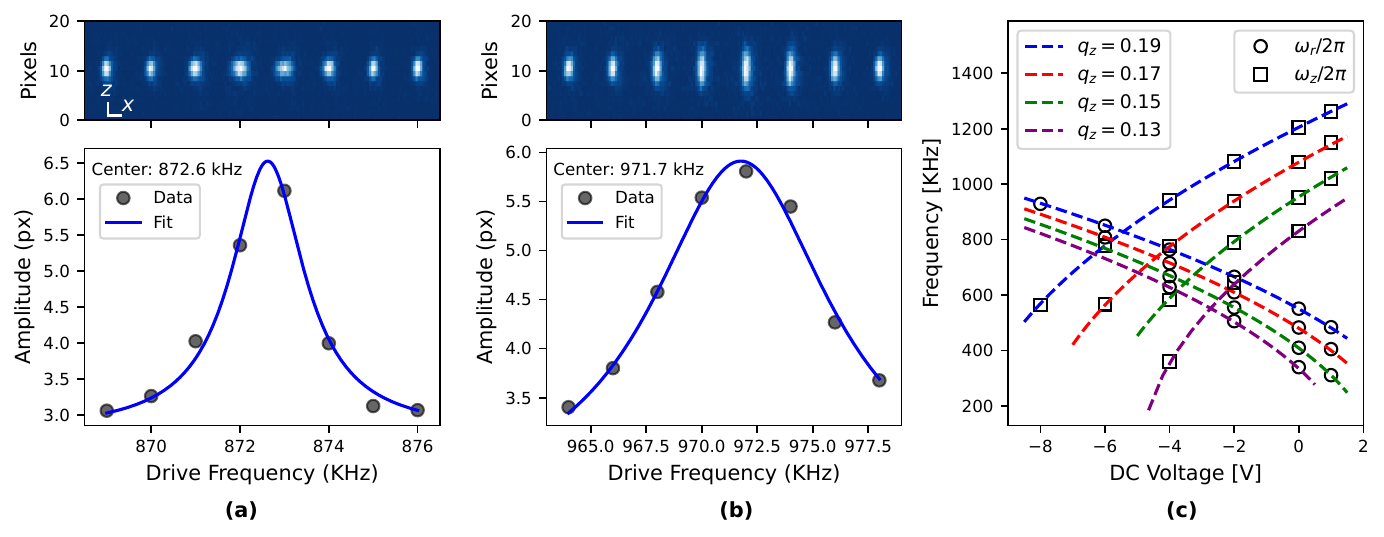}
    \caption{(a) Stitched images of a single ion observed on the EMCCD camera at different frequencies of the time-varying voltage about the resonance. As the frequency approaches the resonance, the image of the ion begins to spread along the $x-$axis. The trap drive voltage amplitude is  234 V, and the applied dc offset is $-5.25$ V. (b) Stitched ion images showing the spread of the ion along the $z-$axis as frequency approaches the resonance. The trapping conditions are the same as in (a). (c) Plot showing the radial and axial secular frequencies for different trap configurations. Experimentally observed data points are denoted by discrete points, and the dashed lines represent the corresponding fits.}
    \label{fig:secular-frequency}
\end{figure}

Experimental determination of the quadrupole coefficient, $A_{2}$, is necessary to compare the fabricated trap model against the designed one. The forced oscillation method \cite{landauMechanics1976} is used to measure the secular frequency of a single ion in the axial and radial directions. In this method, the trapped ion is driven by a small time-varying voltage, and the frequency of the applied voltage is scanned. The energy gain is maximum when the drive frequency matches the system's natural frequency and fluorescence from the ion drops, as shown in Figure \ref{fig:secular-frequency}.

In the adiabatic approximation, the relation between the drive frequency, secular frequency, and Mathieu parameters has the following form \cite{majorChargedParticleTraps2005}  
\begin{equation}
    \omega = \frac{\Omega}{2}\sqrt{a+q^{2}/2} , \label{eq-secularfrequency} 
\end{equation}
where $\omega$ is the secular frequency, $a$ and $q$ are dimensionless Mathieu parameters and $\Omega$ is the drive frequency. For modified trap geometries, the relation between the trap parameters and the Mathieu $a$ and $q$ parameters is given by the following equation 
\begin{equation}
\eqalign{
    2a_{x} =  2a_{y} = -a_{z} &= \frac{8A_{2}Q U}{m z_{0}^{2}\Omega^2}, \cr
   -2q_{x} = -2q_{y} = q_{z} &= \frac{4A_{2}Q V}{m z_{0}^{2}\Omega^2}, 
}
\label{eq-a-q-relation}
\end{equation}
where $Q$ is the charge of a single ion, $U$ is the dc voltage applied to the trap, $V$ is the rf voltage applied to the trap, and $m$ is the mass of a single trapped ion. Equation \ref{eq-secularfrequency} and Equation \ref{eq-a-q-relation} are used to fit the secular frequencies against the given trap parameters to find the quadrupole coefficient. Experimental data points and fit used to determine the quadrupole coefficient are shown in Figure \ref{fig:secular-frequency}. New variables are introduced in the equation used for fitting to account for the error in the probe voltage and an overall dc offset. The value of the fitted parameters is mentioned in Table \ref{fitted-parameters}. Errors mentioned in the fitted parameters refer to the standard deviations associated with the values of the fitted parameters.

\begin{table}[b]
\caption{Fitted trap parameters}
\label{fitted-parameters}
\footnotesize
    \begin{tabular}{@{}p{3cm}p{3cm}p{3cm}p{4cm}@{}}
    \br
     Parameters & $A_{2}$ & dc offset error (V) & Probe voltage error ($\%$)\\
    \mr       
    Design & $0.298 \pm 0.001$ & $0$ & $\approx 1 $\\
    Fitted  & $0.300 \pm 0.002$  & $0.571\pm0.016$ & $\approx 4.7$\\
    \br 
    \end{tabular}\\
\end{table}
\normalsize

The experimentally determined value of the quadrupole coefficient is $0.300\pm 0.002$, which agrees with the designed parameters. A significant variation in the probe voltage error can be attributed to changes in the characterization environment. The probe is characterized at 16 dBm, and in the setup, 40 dBm is applied across it. The change in power dissipation across the probe can affect its value, which in turn can alter the sampled voltage.

\subsection{Excess Micromotion compensation}

\begin{figure}
    \centering
    \includegraphics[width=0.7\linewidth]{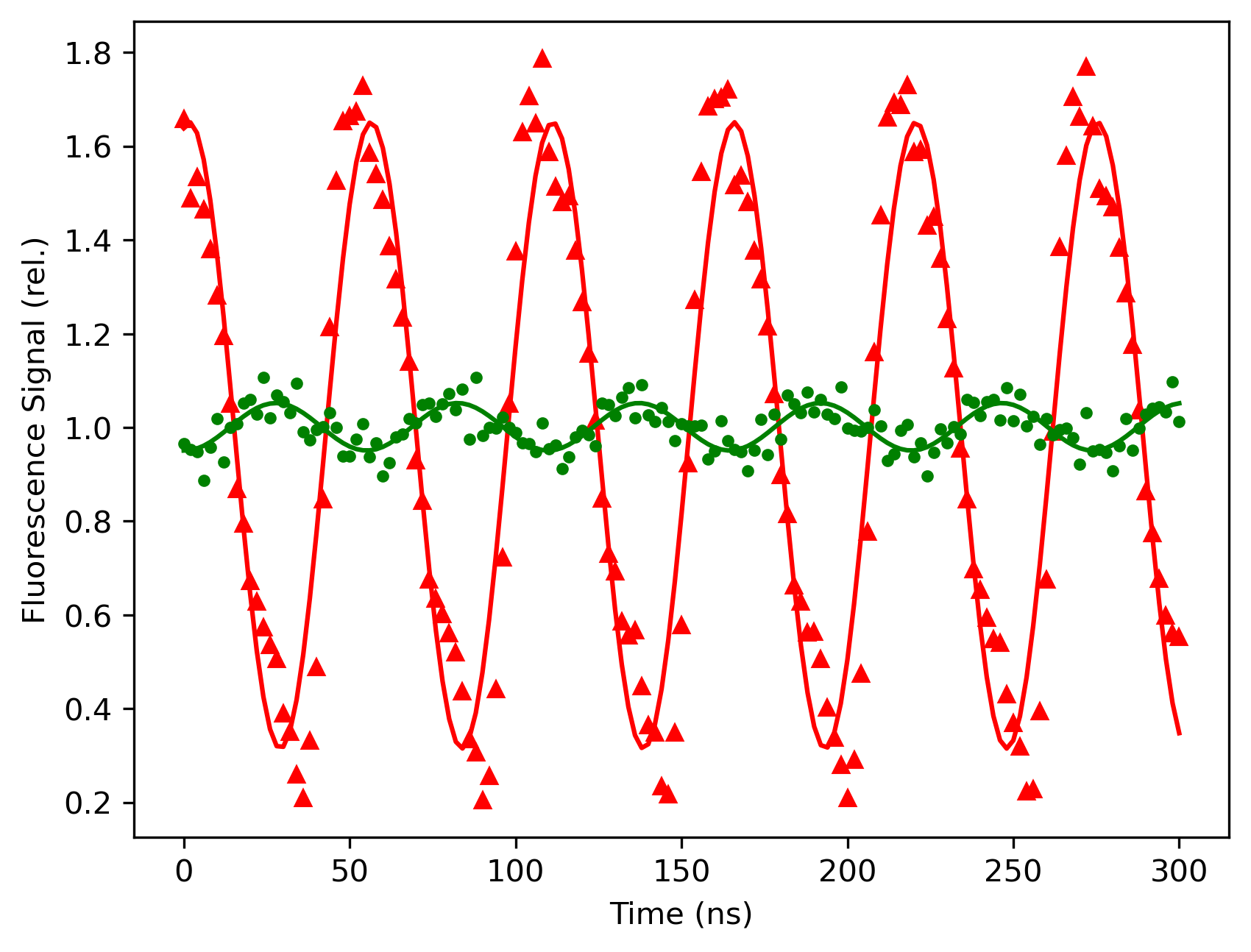}
    \caption{Red and green points are the photon correlation histogram due to the EMM along the spectroscopy beam before and after compensation, respectively. The red and green lines are the fits corresponding to them.}
    \label{fig:micromotion}
\end{figure}

Micromotion is an inherent property of an ion trapped in a Paul trap. It can be decreased by cooling, but cannot be completely eliminated. EMM arises when the null points of the rf and stray dc fields do not coincide at a single point. This results in driven motion of the ion, so cooling cannot eliminate it. The photon-correlation method \cite{berkelandMinimizationIonMicromotion1998,kellerPreciseDeterminationMicromotion2015} is used to determine the EMM. The driven motion has a fixed phase relation with the trap drive voltage, resulting in a correlation between the fluorescence modulation due to EMM and the trap drive voltage. The histogram of the time delay between the arrival of a photon and the subsequent zero crossing of the RF voltage in the negative direction has the following form
\begin{equation}
    S = S_{0} + \Delta S \cos(\Omega t- \phi), \label{eq-correlation} 
\end{equation}
where $S_{0}$ is the mean fluorescence, $\Delta S/S$ and $\phi$ are the amplitude and phase of the correlation signal, respectively. Three non-parallel and non-planar beams are aligned to sample the micromotion along all three directions. First, the correlation signal for individual beams is measured for uncompensated cases and compensated by manually changing the compensation and outer electrode voltages. The correlation signal for the uncompensated and compensated cases along the spectroscopy beam is shown in Figure \ref{fig:micromotion}. Along the spectroscopy beam, $\Delta S/S_{0}$ is minimized to $\approx 0.05$, which corresponds to a relative frequency shift due to EMM at the level of $3.5\times 10^{-18}$ for $^{40}$Ca$^{+}$.

\subsection{Structural Transitions of few-body Coulomb Crystals}
The capabilities of the trap extend beyond single-ion spectroscopy to the study of mesoscopic few-body physics of laser-cooled Coulomb clusters. By varying the dc voltage applied to the outer electrodes ($U_{\mathrm{dc}}$), we can continuously tune the trap anisotropy parameter $\alpha = \omega_z / \omega_x$, where $\omega_z$ and $\omega_x$ are the axial and radial secular frequencies, respectively. This control allows us to deterministically prepare $N$-ion Coulomb crystals in distinct 1D and 2D configurations, effectively creating ``test targets" with tunable inter-ion spacings to verify the performance of our imaging system.

\begin{figure}
    \centering
    \includegraphics[width=\linewidth]{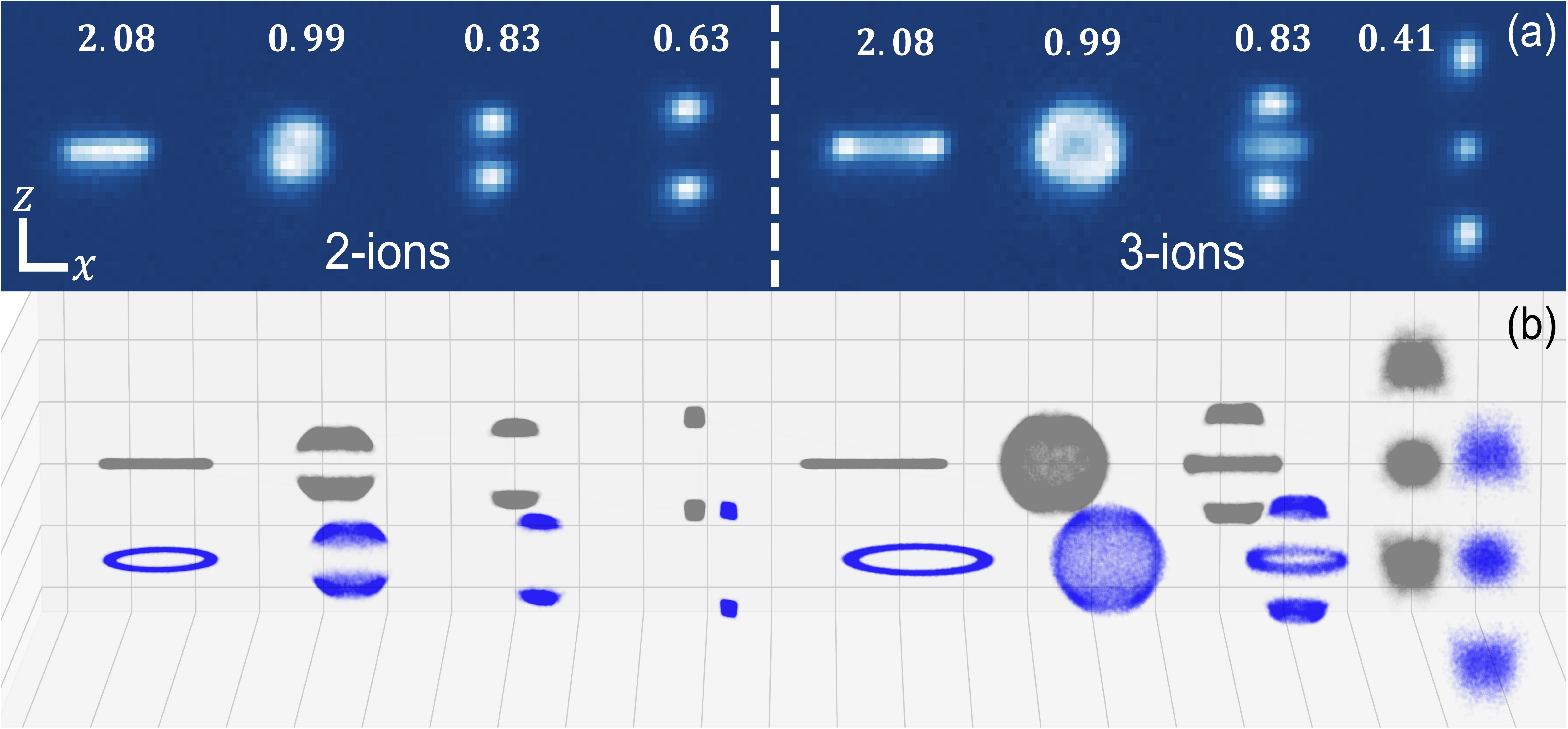}
    \caption{(\textbf{a}) Stitched fluorescence images of $2$ and $3$-ion clusters, imaged from the experiments at different trap anisotropies $\alpha$ displayed above. For scale, the bars on the bottom left represent $5\; \mu$m $\times$ $5\; \mu$m. (\textbf{b}) The 3D scatter plots from MD simulations performed at the corresponding anisotropies, including stochastic photon kicks. Their projections onto the \textit{x-z} plane are shown in grey and are in excellent agreement with the experimental images. The grid lines are spaced at $5\; \mu$m $\times$ $5\; \mu$m. The smearing is due to a Goldstone mode corresponding to the free rotation about the \textit{z}-axis.}
    \label{fig:2+3ions}
\end{figure}

We investigated the structural transitions of $N=2$ and $N=3$ ion clusters, where the configurations are determined by the competition between the external confinement and the mutual Coulomb repulsion. To obtain the configurations numerically, we employed Molecular Dynamics (MD) simulations that integrate the equations of motion under the full time-varying trap potential $\phi(r, z) \cos(\Omega t)$ and explicitly incorporate stochastic momentum recoils from photon scattering events to model the thermodynamics of Doppler cooling \cite{knoop_numerical_2016}. For a 2-ion crystal in the prolate trapping potential ($\alpha < 1$), where radial confinement dominates, the ions form a ``dumbbell" along the trap axis ($z$). As the anisotropy is tuned to the oblate regime ($\alpha > 1$), the ``dumbbell" structure is oriented perpendicular to the trap axis in the $x$-$y$ plane. Due to the cylindrical symmetry of the trap ($\omega_x \approx \omega_y$), this planar configuration is free to rotate about the $z$-axis (Goldstone mode), appearing as a ring within the integration time of fluorescence images. 

For the $N=3$ cluster, a richer set of transitions is observed. At low anisotropy ($\alpha < 0.64$), the ions form a linear chain. As $\alpha$ is increased, the transverse confinement squeezes the chain, leading to a displacive transition at a critical value of $\alpha_c \approx 0.64$. At this point, the linear chain buckles into an isosceles triangle in a plane containing the $z$-axis. As $\alpha$ is further increased to unity ($\alpha = 1$), the symmetry of the potential becomes isotropic, and the cluster transitions into an equilateral triangle for $\alpha\geq1$, which again settles into \textit{x-y} plane for $\alpha > 1$. These observations demonstrate the trap's ability to precisely control the potential landscape, a prerequisite for studying complex symmetry-breaking dynamics in larger clusters \cite{ayyadevaraObservingDynamicsOctupolar2025a, ayyadevaraSymmetrycontrolledThermalActivation2026}. Moreover, the ability to spatially resolve the individual ions within the observed configurations, where the MD-predicted separation is $\sim 5 \mu$m, directly demonstrates the performance of our custom lens-assembly.

\section{Conclusion}
We have presented the design, fabrication, and characterization of an endcap-type Paul trap.   Controlled loading of one or more ions is demonstrated. The experimentally determined value of the quadrupole coefficient, $0.300\pm 0.002$, is in excellent agreement with the design value of $0.298\pm 0.001$. We demonstrate micromotion minimization,  which corresponds to a relative frequency shift of $3.5 \times 10^{-18}$ for $^{40}$Ca$^{+}$. Going forward, the EMM can be minimized in 3D with automation. Moreover, the relative frequency shift due to EMM can be further eliminated by operating the trap at the ``magic frequency" \cite{huangCa40Ion2019a} where the second-order Doppler shift and the Stark shift due to the ac field along the spectroscopy beam cancel each other. 

We also demonstrate a controlled study of $2$ and $3$-ion Coulomb clusters, which are very well resolved by our custom imaging system. The planar structures of these clusters undergo free rotation about the trap axis, illustrating the near-perfect cylindrical symmetry of the trap potential. The excellent agreement between the numerical simulations and the observed cluster configurations, beyond $2$ and $3$-ions has been reported in a separate work \cite{ayyadevaraObservingDynamicsOctupolar2025a}, where the dynamics of octupolar structural transitions in $4$, $5$, and $6$-ion clusters were studied. In addition, owing to precise control over the trap anisotropy $\alpha=\omega_z / \omega_x$, thermally activated dynamics of the $5$-ion pyramidal cluster have been used to test the multidimensional Kramers-Langer theory of reaction rates \cite{ayyadevaraSymmetrycontrolledThermalActivation2026}. These results illustrate that the trap's performance is optimal, enabling both precision spectroscopy and studies of the mesoscopic physics of Coulomb clusters.

\ack

We thank GTTC Rajajinagar for machining the electrodes, CMTI Peenya, Josh Hacko(Sydney Watches) for fruitful discussions, and RRI Workshop for machine design and overseeing the fabrication process. We thank Harikrishna Sahu, Md Arsalan Ashraf, Nishant Joshi, Somshekhar S, Keerthipriya S, Sujatha, and Meena M S from RRI for their assistance during the design and assembly of the experimental setup, and Sankalpa Banerjee for designing the helical resonator.

We thank Piet Schmidt, Christian Tamm, Nils Huntemann, Tanja Mehlst\"aubler, and  Martina Knoop for technical assistance; Vashudha K N, Cell Bio-Physics lab, Bio-Physics lab, and Rheo DLS lab from RRI for sharing equipment;  Shanti Bhattacharya for assistance with Zemax. 

The authors thank Indian Science Technology and Engineering facilities Map (I-STEM), a Program supported by Office of the Principal Scientific Adviser to the Govt. of India, for enabling access to the COMSOL Multiphysics 5.6 software suite used to carry out this work.

The authors acknowledge support from the Ministry of Electronics \& Information Technology(MeitY), Government of India, under the “Centre for Excellence in Quantum Technologies” grant with Ref. No. 4(7)/2020-ITEA and QuEST program of DST, IUCAA. EK thanks the Department of Atomic Energy, India, for the Raja Ramanna Fellowship.

\section*{References}

\providecommand{\newblock}{}

\end{document}